\documentstyle[preprint,aps,floats,epsfig]{revtex}
\tighten
\begin{document}
\draft
\preprint{MIT-CTP-3331}
\topmargin=0.1cm

\title{From boom to bust and back again: the complex dynamics of trends and 
fashions}
\author{Lu\'{\i}s M. A. Bettencourt}
\address{Center for Theoretical Physics, 6-316,  
Massachusetts Institute of Technology,
Cambridge MA 02139}

\date{\today}
\maketitle

\begin{abstract}
Social trends or fashions are spontaneous collective decisions made by 
large portions of a community, often without an apparent good reason.
The spontaneous formation of trends provides a well documented mechanism 
for the spread of information across a population,  the creation of culture  
and the self-regulation of social behavior. 
Here I introduce an agent based dynamical model that captures the essence 
of trend formation and collapse. The resulting population dynamics alternates 
states of great diversity (large configurational entropy) with the dominance 
by a few trends. This behavior displays a kind of self-organized criticality, 
measurable through cumulants analogous to those  used to study percolation. 
I also analyze the robustness of trend dynamics subject to external 
influences, such as population growth or contraction and in the presence of 
explicit information biases. The resulting population response gives 
insights about the fragility of public opinion in specific circumstances 
and suggests how it may be driven to produce social consensus or dissonance.
\end{abstract}

\pacs{PACS Numbers: 89.75.Fb, 89.75.Da, 89.65.Ef, 89.65.Gh}


\section{Introduction}
\label{sec1}

Making choices about which social circles to join or evade is one 
of the most ubiquitous, important and difficult decisions facing each one 
of us every day. The basis for this difficulty is that our social environment 
is too complex for us to predict the detailed outcome of our actions.
It is also a dynamical environment so that past experience may
be a poor indicator of future events. 
As a result many of our most important choices must 
be made within a limited timespan and without full information. 

These practical limitations make many different courses of action 
seem equally viable. This degeneracy can be described 
mathematically as a symmetry among all equally good choices at 
the individual level. This {\it choice degeneracy} is also familiar from 
information saturated environments, where useful information is not 
easily discriminated from noise, of times when information is accessible 
but not reliable, or may simply result from a choice not to form our 
own opinions. 

How do we actually make choices in these difficult environments? 
In many cases we rely on the {\it actions} of others we know as the basis 
for our decisions. If their choices appear successful to us we may 
adopt them as our own.
Whether this is a good or bad strategy depends then on 
how well informed our acquaintances are. 
In any case we are  guaranteed not to do worse than most of the people  
that surround us, which may be all that matters. 
In this light the spontaneous emergence of collective behavior corresponds 
to a particular choice, made simultaneously by many agents, 
among others that are equally good 
- a spontaneous breaking of individual {\it choice symmetry} 
by the state of the population as a whole. 

The strategy to base our decisions on the actions of others  
is very universal. The natural languages, for example, 
are rich in related aphorisms. Recently this decision 
making strategy has become the focus of a sizable literature 
in economics \cite{Keynes} and the social sciences \cite{Social}. 
Bikhchandani, Hirshleifer and Welch 
\cite{hirshleifer} (BHW) were, to my knowledge, the first to stress 
the universal nature of trends and fashions \cite{BHWrefs}. 
They also proposed a simple model of sequential choice in which 
they can develop \cite{hirshleifer,earlyherd}. 
BHW called the widespread adoption of a particular 
decision an {\it information cascade}.  The term refers to how a piece 
of information can spread quickly through the whole population 
as a trend. They also collected a vast literature, across many scientific 
disciplines in support of trend dynamics as an important social mechanism 
\cite{Social,BHWrefs}. 

In financial markets for example, where collecting quantitative data 
{\it a posteriori} is relatively easy, analysts \cite{Analysts} and mutual 
fund managers \cite{FundManagers} are observed to follow each other's choices
and recommendations.
In elections it is well documented that opinion poles influence the 
decision process \cite{Voting}. 
Cycles of innovation in industrial production 
\cite{IndustrialInnovation} are determined in part by 
the success of one's industrial peers and/or by the 
fear of lagging behind or losing market share.   
Television programming displays similar patters \cite{TVfads}.
The incidence of crime or fraud \cite{Crime}, e.g. tax evasion, depends in 
part on the observation of others who may have gotten away with it.

There has also been a growing interest in related ideas 
\cite{StatPhys,sneppen} 
in the literature of complex systems and critical phenomena. 
Close relatives of trend dynamics are models of flocking \cite{flocks}
and {\it herd behavior} \cite{earlyherd,herd,herd2}. The latter have been 
developed to model financial markets, 
the former to describe how groups of animals may coordinate their 
movements to form a flock and, more generally, the collective 
dynamics of self-propelled particles. 
Flocking requires the spontaneous breaking of rotational 
symmetry, which happens when the velocity of most moving agents aligns 
to form the familiar pattern of organized collective motion.

In financial models of herd behavior agents are grouped together by 
given rules. Each of these 
groups or herds then makes collective decisions (buy/sell) 
in unison. Trend dynamics 
is somewhat different: each agent actually tries to stay ahead of the crowd,
trading his position for another only if the latter offers greater promise. 
As we shall see below trend dynamics leads to {\it spontaneous} 
consensus even as all agents exercise free will. 
In this sense the class of models considered here will differ from those 
describing herd behavior, although some aspects of the emergent collective 
behavior may be similar.   

Trend setting and trend following can often be thought of as a game,
to be played many times by a large number of participants.
For maximal advantage one would like to join a winning trend at the
earliest possible opportunity, ride it to the height of its popularity, 
and leave before it collapses. 
However, because this strategy is shared by all players, 
it leads to some choices becoming very widespread. 
Once it is apparent that the same state is shared by all social advantage 
is lost and some agents are tempted to leave. At this point dominant trends 
become unstable to decay due to competition from new faster movements. 

Thus we see qualitatively the two most important features of trend dynamics:
the faster a trend is growing the greater its promise i.e. the 
more attractive it feels. Consequently, when a trend becomes widespread and 
cannot sustain its pace of growth, it looses its appeal. Then agents 
begin to look elsewhere for the next 'hot thing'. The purpose of the 
present paper is to capture the essence of trend formation 
and decay, starting at the individual level, in a population facing many 
competing choices. To do that I will construct simple models 
of agents interacting with each other and acting according to 
their (perceived) {\it individual} best interest.

Beyond analyzing the genesis of trends  I will show that social behavior 
under trend dynamics is a prototypical complex system. A unifying property 
common to many complex systems is the interesting way in which their dynamics 
samples large dimensional configuration spaces. Here too we will see that 
the population is driven back and forth between states of order, or large 
configurational information, and disorder. 
Both these extremes are unstable so that the system spends a long  
time in between. This cycling between order and disorder, once time averaged, 
shows interesting analogies with ensemble averages of statistical systems 
in the vicinity of a critical point. Thus trend dynamics can display 
a specific sort of self-organized criticality \cite{Bak}. 

Another interesting aspect of trend dynamics is its susceptibility to 
external influences. Because, as we shall see below, the system has no 
global stable fixed points its dynamics are extremely sensitive to the 
introduction or subtraction of new agents with particular preferences. 
These properties produce insights into the fragility or 
robustness of 'public opinion' in specific circumstances 
and tell us how it may be driven to generate social consensus or dissonance.

The remaining of this paper is organized as follows. 
In sec.~\ref{sec2} I describe the basic agent based model and several 
of its possible variations. In sec.~\ref{sec3} I construct quantities 
that give a global characterization of the dynamics. Firstly I will 
define an information (Shannon) entropy over population configurations. 
Secondly it is natural to define cumulants analogous 
to a percolation strength and susceptibility \cite{percolation}. 
These quantities allow a comparison between time averages of population 
configurations and typical properties of critical phenomena. 
Section~\ref{sec4} is devoted 
to a more detailed analysis of the dynamics. There I show how several 
macroscopic properties can be understood by simpler analytical 
expressions, without requiring knowledge of the whole population.  
Sec.~\ref{sec5} is devoted to some of the properties of the driven system 
where individuals are added to or subtracted from the population with 
specific trend preferences. This will show how trend dynamics can be driven 
by external influences. Finally in Sec.~\ref{sec6} I summarize the results 
and discuss other related problems and applications.

\section{The model and its global dynamics}  
\label{sec2}

In this section I define the agent based model used in the remaining of 
this paper. I will also give a first characterization of the resulting 
dynamics by constructing quantities that capture their most important 
global properties. 

\subsection{Definition of the model and possible variants}

I consider a population with $N$ agents and $L$ {\it trends} or {\it labels}. 
Every agent is characterized by a single label, denoting the group 
or trend he/she belongs to at a given time. 
All labels are equally good from the point of view of an isolated agent. 
For a discrete set of $L$ labels this is a $Z_L$ symmetry.
This prescription implements the basic individual choice degeneracy 
discussed in the Introduction. 
Nevertheless, as we shall see, spontaneous collective choices will be 
generated and subsequently destroyed dynamically. 
The specification of a single label per agent 
can be easily relaxed, but would lead to more complicated 
(and arbitrary) dynamical rules.

At each time every agent $i$ contacts another $j$ at random in his social 
circle. He compares the relative growth (or momentum) 
of their labels $p_i(t)= \Delta N_i(t)/N_i(t)$, where $N_i(t)$ is the number 
of agents in label $i$ at time $t$ and $\Delta N_i(t)=N_i(t)- N_i(t-1)$. 
If his label's momentum $p_i$ is smaller that $p_j$ he adopts 
the other agent's trend; otherwise he keeps his \cite{function}.

Moreover if the momentum of a given trend is smaller than a given 
value $p_{\rm crit}$, agents leave their label for a brand new (empty) 
one. Thus $p_{\rm crit}$ parameterizes conformism; the threshold momentum 
below which staying in a slow moving trend becomes unbearable and taking 
the risk on a brand new thing is preferred. 
This parameter will play an important role in trend dynamics, 
as we shall see below. Here, I will assume, for simplicity, that all 
agents share the same value of $p_{\rm crit}$.

A few extra remarks are in order:

i) the values of $N$ and $L$ can be made functions of time, 
reflecting population growth or decrease and the change in their 
realm of choices. Some of these scenarios will be explored in 
sec.~\ref{sec5}.

ii) Each agent's social circle can be limited to subsets of the population;
as it happens in reality in social networks \cite{SocialNReview}. 
These restrictions effectively condition the flow of information and 
make the choices of certain individuals more influential that those of others. 
These properties lead to fascinating dynamical effects, which
can depend sensitively on the morphology of the network. 
Such issues require an adequate discussion of social networks and 
a parametric exploration of some of their structural freedoms. 
For this reason the influence of the structure of social networks on the 
dynamics of trends will be presented in a separate publication \cite{bett2}.
Below I shall use an unstructured population, where each agent can contact 
any other with equal probability.

iii) The individual choice to change label can be made according to a more 
sophisticated probabilistic transition amplitude. 
This becomes essential if more parameters condition the choice 
or if agents do not act deterministically on the facts. 
Although these refinements may be necessary to model complex situations 
I shall not consider them in the context of the present paper.

iv) The value of $p_{\rm crit}$ is taken below as an input, 
a property shared by all agents. An interesting possibility is that its 
value may reflect a global sentiment that can 
change in time with the state of the population. 
Then $p_{\rm crit}$  should be computed self-consistently in time, 
e.g. conformism may decrease as agents perceive that most others 
belong to their trend or may rise during difficult times of population 
decrease. 

\begin{figure}
\begin{center}
\epsfig{figure=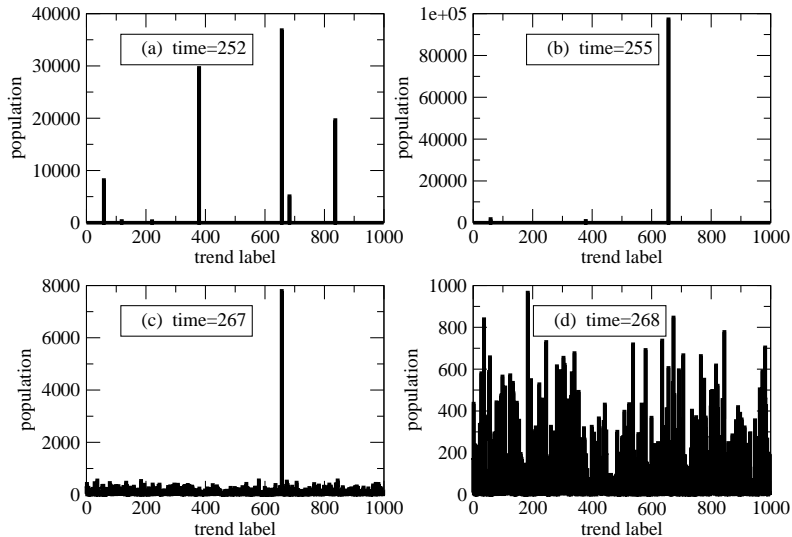,width=3.6in,angle=270}
\caption{Snapshots of trend occupation numbers
for an evolution with $N=10^5$, $L=10^3$ and  $p_{\rm crit} =10^{-5}$.
The system cycles between ordered states, characterized by the dominance
of a few trends (a)-(b), their decay (c), and the emergence of many small 
competing labels (d), from which a few dominant ones emerge again and so 
on.}   
\label{fig1}
\end{center}
\end{figure}

Fig.~\ref{fig1} shows a typical evolution for a low value of 
$p_{\rm crit}=10^{-5}$. Clearly dominant trends alternate with periods of 
coexistence of many competing labels from which one eventually emerges to 
dominate and so on. The spontaneous formation of a large trend corresponds 
qualitatively to the phenomenon that BHW coined an 
{\it information cascade} as a label becomes widespread in the whole population.
Here, however, the realm of choices is arbitrarily large and
the population is not organized in a queue making sequential choices. 
This will allow us richer dynamics and closer similarities to systems 
familiar from statistical mechanics.
Moreover the agents in the present model do not know about the previous 
choices of all others before them; they can merely compare the progress 
of their labels to those of their neighbors. 
Technically the implementation used here is a Markov chain, where the state 
of the system at one time (its occupation numbers and their momenta) is 
determined (stochastically) from its state at the previous time. In this 
sense agents are not aware that they may be joining a large scale movement, 
they are just searching for the most promising choice in their realm of 
observation. The degeneracy of choice, built in at the individual level, 
persists as collective movements emerge dynamically because the winning 
trend can assume any of the $L$ labels \cite{Movies} with equal probability.

The dynamical structure of the system implies that there is no global stable 
fixed point. Demanding that a configuration be static 
implies $p_i=p_j=p$. If we insist that all $p_i=p$ are the same 
then $p=0$ because of the overall conservation of individuals. 
This is a trivial static point. It can be realized in 
many ways (any arrangement of $N$ individuals in $L$ labels); 
most of which are close to the flat distribution, the most 
disordered state of the population.  
This fixed point is clearly unstable to any 
perturbation: if a single trend acquires positive momentum 
(and another negative from agent number conservation) 
the system will move away from the original configuration 
as all agents try to join the former and exit the latter. 

A very particular instance of this fixed point is the asymptotic situation 
where $N_i \rightarrow N$, $p_i \rightarrow 0^+$ and $N_j \rightarrow 0$, 
$p_j\rightarrow 0^-$ $\forall_{j\neq i}$, 
which corresponds to the growth of a single dominant trend
at the expense of all others. 
This situation can be realized for each one of the $L$ labels.
A single fully occupied trend is the most ordered state the system can take. 
For it not to be {\it stable} it is necessary that 
$p_{\rm crit} > 0$ - this is the first crucial role of $p_{\rm crit}$.
Otherwise, if $p_{\rm crit} \leq 0$, agents have no desire to leave a
static (or decaying) trend in the absence of faster growing competitors and 
the evolution freezes when the first dominant trend is formed.

\subsection{Global characterization of trend dynamics}

As we have already seen one of the features of the evolution at 
small $p_{\rm crit}$ is the formation of widespread trends, 
i.e. situations where one of the labels is 
substantially more populated than all the others.  
However, as it does so, the dominant trend must slow down and 
will eventually collapse into many small, fast moving trends, which 
proceed to compete for dominance, see Fig.~\ref{fig1}. 
The instability of these states 
leads to characteristic cyclic (but aperiodic) dynamics. 
This section is dedicated to constructing global quantities 
inspired by analogies to statistical physics that capture these properties.

First, the qualitative sense that the population is alternatively in states 
of disorder (when many trends coexist) and order 
(where one emerges as dominant) can be captured by defining a Shannon 
entropy $S$, over the discrete set of labels 
\begin{eqnarray}
S = - \sum_{i=1}^L  n_i \ln n_i,
\end{eqnarray}
where $0\leq n_i \leq 1$ is the probability of finding an individual 
in label $i$. Thus by knowing $n_i$ as a function of time we can measure 
the evolution of the total entropy of the system. $S$ is not conserved 
because the evolution is not Hamiltonian.

\begin{figure}
\begin{center}
\epsfig{figure=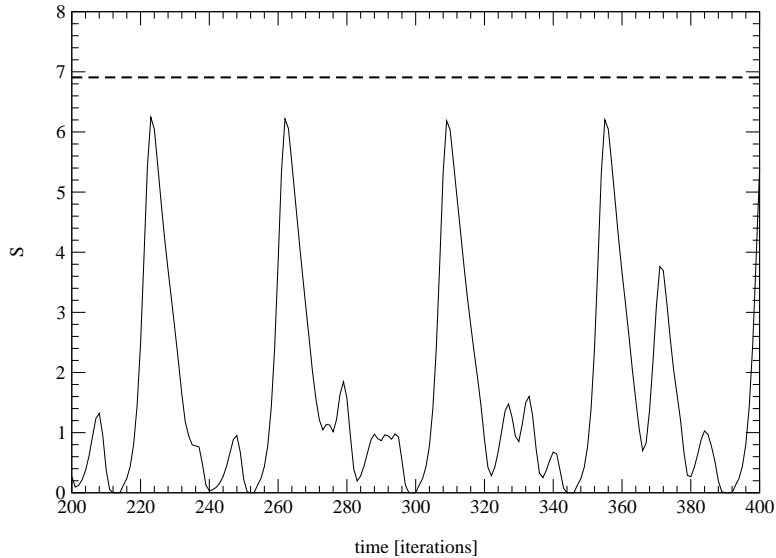,width=3.6in,angle=270}
\caption{The behavior of the total entropy S, through several cycles 
of trend formation and decay, for a population with $N=10^5$ 
agents, $L=10^3$ labels and $p_{\rm crit}=10^{-5}$. 
The dashed line shows the maximal entropy 
$S_{\rm max} = \ln(L) \simeq 6.91$, corresponding to the flat distribution.}  
\label{fig2}
\end{center}
\end{figure}

The value of $n_i$ leading to the highest entropy $S_{\rm max}$ is the 
flat distribution 
\begin{eqnarray}
n_i={1 \over L}, \ \forall_{i}.
\end{eqnarray}
Then the entropy $S_{\rm max}=\ln(L)$. 
In contrast for one single trend containing the whole population 
$S_{\rm min}=0$. Dynamically we expect the system to alternate between 
these two asymptotic states, at least at low $p_{\rm crit}$. 
Fig.~\ref{fig2} shows several cycles of growth and decay in the entropy $S$, 
for $p_{\rm crit}=10^{-5}$, $N=10^5$ and $L=10^3$. 

We have already anticipated that the dynamics of trends, in analogy with  
many other complex systems, may display certain forms of self-organized 
criticality. I now define quantities that allow us to diagnose such behavior. 
The simplest and most paradigmatic critical phenomenon is percolation 
\cite{percolation}. 
As in other second order transitions the critical point is associated with the 
the divergence of a characteristic macroscopic length, together 
with several other susceptibilities, associated with the system's response to 
macroscopic stresses.
	
Statistical distributions in the vicinity of the critical point can be defined 
by two critical exponents (and by dimensionality).  
For percolation  these two exponents can be obtained 
from the behavior of two independent ensemble averages, usually the 
percolation strength $P_c$ (the "size" of the largest 
correlated cluster) and a percolation susceptibility $S_c$, defined by  
the sum of squares of the cluster sizes with the largest cluster subtracted. 
Criticality is the onset of the formation of a spanning cluster and 
is associated with a sharp increase of $P_c$ and a peak (in the finite volume)
in $S_c$. In the thermodynamic limit, where the number of degrees of freedom 
tends to infinity (the infinite volume limit), $S_c$ usually diverges with 
some characteristic critical exponent. 
\begin{figure}
\begin{center}
\epsfig{figure=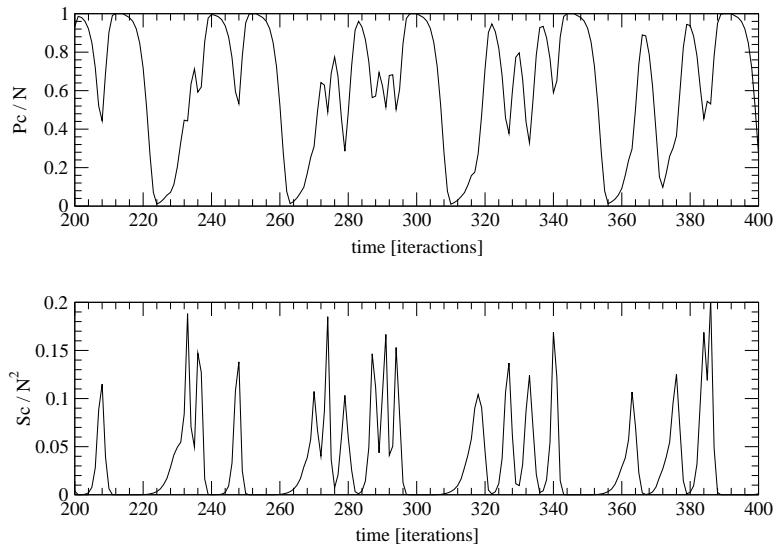,width=3.6in,angle=270}
\caption{The behavior of the percolation strength $P_c$ (upper panel) 
divided by $N$, and the percolation susceptibility $S_c$ (lower panel),
divided by $N^2$, for the evolution of Fig.~\ref{fig1}. 
$P_c$ cycles through states of maximal occupancy 
$P_c=1$ and very low occupancy $P_c \simeq 0$. The intermediate state, 
at the onset of the formation of a dominant term, is characterized 
by a sharp maximum of $S_c$, signaling critical behavior.} 
\label{fig3}
\end{center}
\end{figure}

In the context of our model each agent can exist in one of $L$ states.
Thus a percolation strength $P_c$ can be defined as the fraction of the 
population in the largest trend $P_c = {\rm max}(N_i)$. The percolation 
susceptibility $S_c$ is then defined as 
\begin{eqnarray}
S_c = \left( \sum_{i=1}^L N_i^2 \right) - P_c^2.
\end{eqnarray}
A limit analogous to the thermodynamic limit in statistical 
systems can be taken by letting the number of agents $N$ tend to infinity 
while also increasing the number of labels $L$ such that the ratio 
of agents to labels $N/L$ stays constant. 

\begin{figure}
\begin{center}
\epsfig{figure=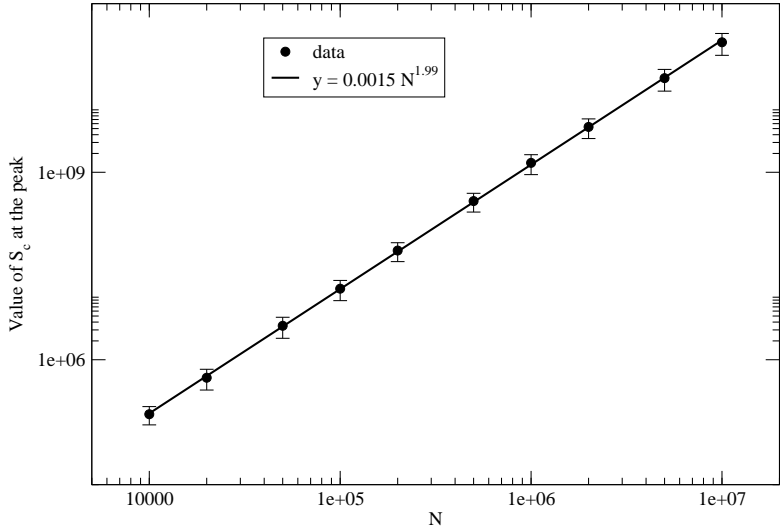,width=3.6in,angle=270}
\caption{The average value of $S_c$ at the peak, as the number of agents
$N\rightarrow \infty$, at fixed $N/L=100$. $S$ diverges 
with $N^\gamma$ with an exponent $\gamma \simeq 2$ (here 
$\gamma=1.99 \pm 0.01$).}
\label{fig4}
\end{center}
\end{figure}

Fig.~\ref{fig3} show the evolution of $P_c$ and $S_c$ for the example of 
Fig.~\ref{fig2}. We see that when a label emerges as the dominant trend 
$P_c$ grows rapidly, and $S_c$ goes through a sharp maximum. 
Fig.~\ref{fig4} shows the variation of the peak of $S_c$ with $N$, 
at fixed $N/L=100$. The behavior of $S_c$'s peak  {\it vs.} $N$ 
shows a divergence as $N \rightarrow \infty$, making a good case 
for the analogy between the dynamical spontaneous symmetry breaking 
of choice symmetry described here and critical phenomena.  

\section{Boom and bust: the rise and fall of trends}
\label{sec3}

As we have seen in the previous section the global motion of the system
can be characterized by a few simple quantities for which we have 
some intuition from statistical physics.
In this section I analyze trend dynamics in more detail and  
derive semi-analytic reduced descriptions for some of their properties.

I start with the decay of a dominant trend. This is a simple 
process because it involves the transfer of agents from the main label 
to all others and, at least initially, can be understood without 
taking into account the detailed dynamics of the latter. 

\begin{figure}
\begin{center}
\epsfig{figure=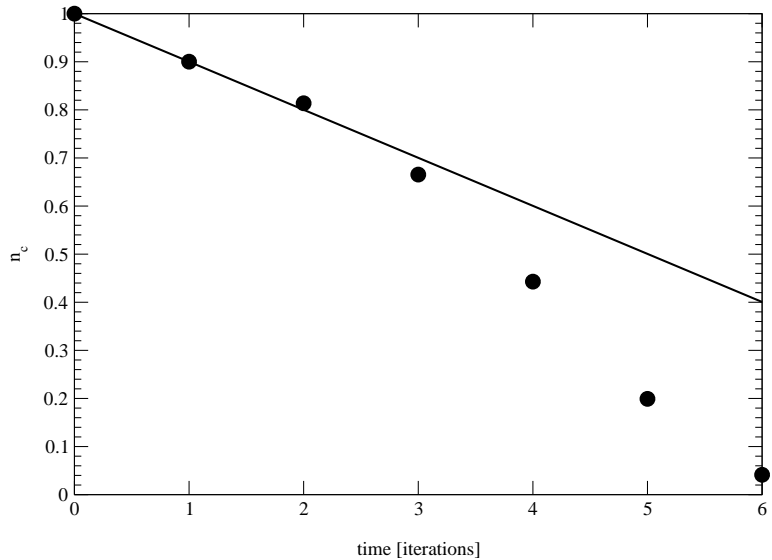,width=3.6in,angle=270}
\caption{The early decay of the dominant trend for an evolution 
with $N=10^4$ and $L=10^3$. Initially the decay proceeds  
by agents spontaneously seeking new, empty labels and is given approximately 
by $n_c(t+1) =n_c(t) - (L-1)$, shown as the solid line (see text).}  
\label{fig5}
\end{center}
\end{figure}

As discussed in sec.~\ref{sec2} the dominant cluster will only decay 
if some individuals choose to leave it, even though there may not be 
any alternative faster growing label available at that time.
For the detailed implementation I choose that the decision to abandon a trend 
is made when its momentum becomes smaller than a certain value 
$p_{\rm crit}$.
When $p_i < p_{\rm crit}$ each agent makes 
a trial search of label space at random;  if an  empty label is found 
it is adopted. Then the first few steps in the decay of 
the main cluster result in $n_c(t+1) =n_c(t) - (L-1)$, see Fig.~\ref{fig5}. 

As labels become filled many small fast growing trends are formed 
and the usual momentum comparison between agents becomes the dominant 
dynamical force.  Then the occupation number distribution 
will be characterized by the dominant trend with negative momentum 
and all other much smaller trends, initially with large positive momentum. 
In these circumstances the decay of the main cluster is dictated by the 
probability of an individual belonging to it to find another individual 
outside. This process it is approximately described by 
$n_c(t+1) = n_c(t) - p_{\rm out} n_c(t)$, where $p_{\rm out}$ 
is the probability of finding an individual {\it outside} the main cluster,  
$p_{\rm out} = {\sum_{k\neq c} n_k}$. 

\begin{figure}
\begin{center}
\epsfig{figure=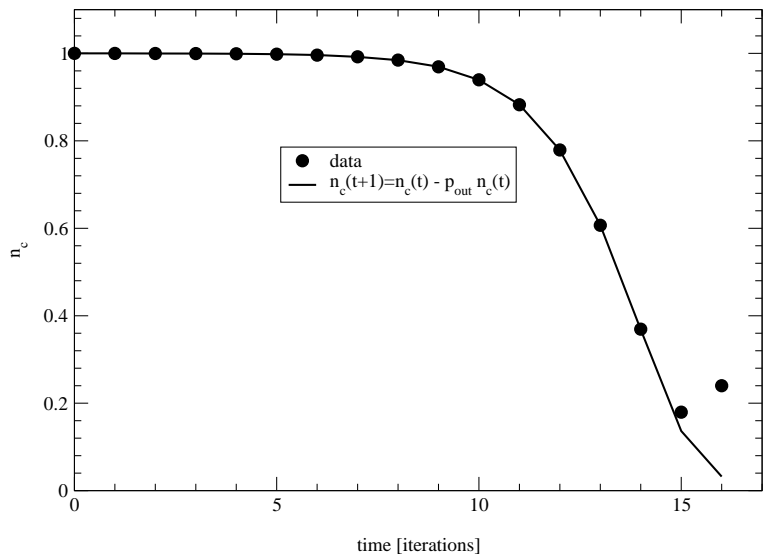,width=3.6in,angle=270}
\caption{The decay of the largest cluster for 
$N=10^5$ agents and $L=10$. The line shows the 
initial prediction given by $n_c(t+1) = n_c(t) -p_{\rm out} n_c(t)$. 
The last point shows a new cluster that has overtaken the former
dominant one in size}  
\label{fig6}
\end{center}
\end{figure}
Figure~\ref{fig6} shows that 
this expectation fits the decay of the main cluster extremely well, even 
at relatively late times. This description breaks down as  other trends
become similar in size to the original dominant one.

\begin{figure}
\begin{center}
\epsfig{figure=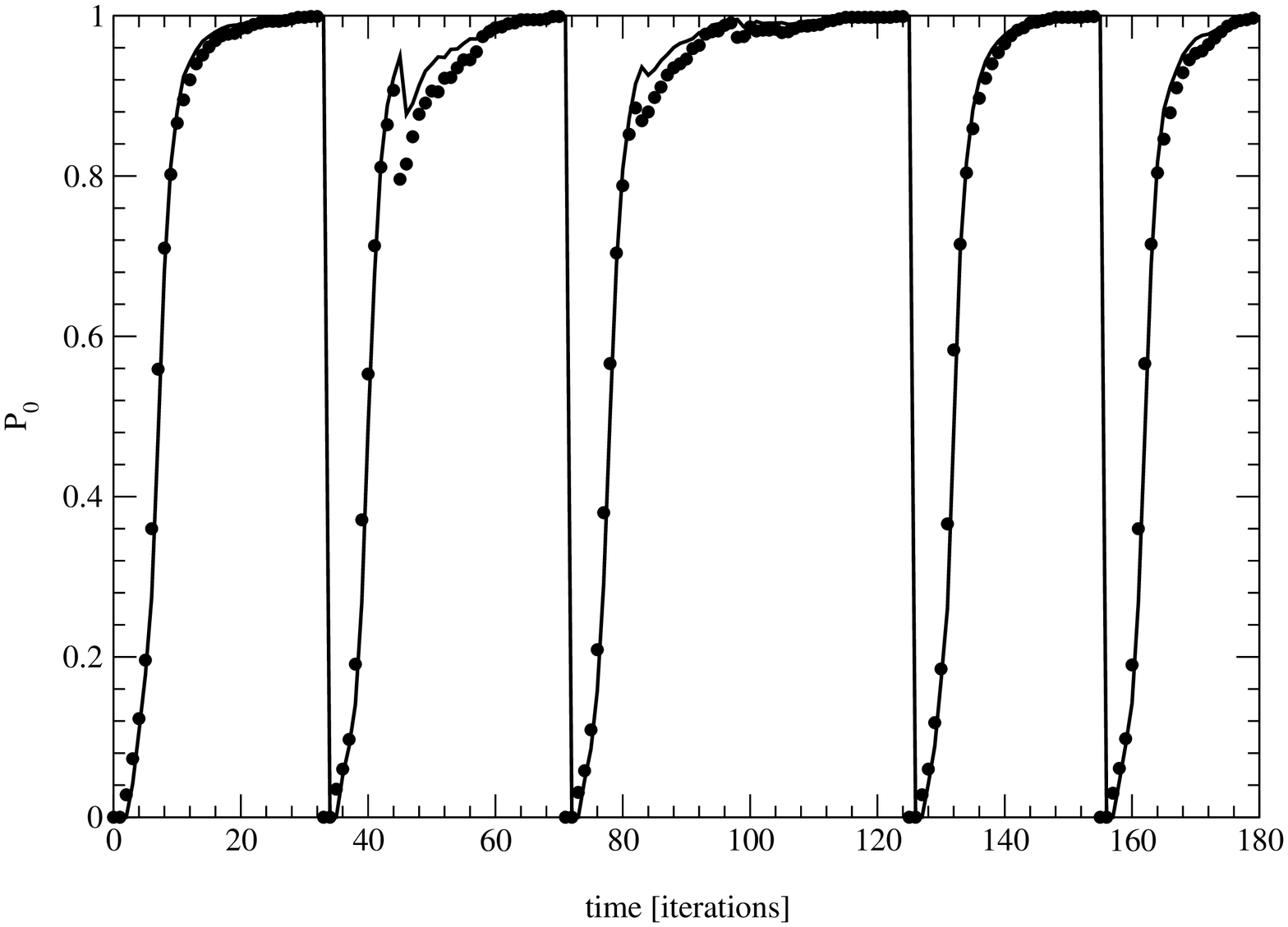,width=3.6in,angle=270}
\caption{The evolution of the probability of a label being empty 
$P_0$ (data points), through several dynamical cycles with $N=10^4$, 
$L=10^3$ and $p_{\rm crit}=10^{-5}$. 
As dominant trends emerge the number of competing labels 
decreases, resulting in an effective dynamical reduction of label choices. 
All labels are subsequently temporarily re-populated as large clusters 
collapse. The evolution of $P_0$ is well described 
by Eq.~(\ref{zeros}) (solid line).}  
\label{fig7}
\end{center}
\end{figure}

The subsequent growth of trends is more difficult to analyze because it 
involves the comparison between  many different competitors.
Note, however, that as dominant trends emerge the system undergoes 
an effective reduction of its phase space. This is because many labels 
become empty. The evolution of the probability to find an empty label
$P_0$ is reasonably well described by  
\begin{eqnarray}
P_0 (t+1) = P_0 (t) + \left[ {1-P_0(t) \over 2} \right] P_0(t),
\label{zeros}
\end{eqnarray}
see fig.~\ref{fig7} for a comparison to data. Eq.~(\ref{zeros}) is based
on the simple expectation that at each time step half of the filled trends
become empty.
Because of this dynamical {\it thinning}  of the number of available 
labels the late evolution of trend formation becomes fairly simple 
as it is characterized only by a few choices.
 
It is useful to consider some very simple cases.
If only two trends are present, a typical {\it endgame} situation,  
the outcome is invariably that the faster growing one always wins, 
even if it is initially smaller. 
In fact, because of the total agent number is conserved they cannot both grow.
This may not always reflect the real world of e.g. brand or political 
party competition, situations that can in many instances be dominated by 
choosing between two alternatives. In reality in these cases the  
relative growth of mutual competitors has become closely monitored.
Intervention in the form of e.g. publicity campaigns is used to change 
agents perceptions leading to more complicated decision making processes, 
beyond the scope of the present model.

If three or more trends are present more interesting situations are 
possible. A typical situation that we explored above is that one of 
the labels is decaying and feeding the growth of others. 
In this situation the decaying cluster functions as a source that 
allows other trends to grow simultaneously. The fate of the latter 
is largely determined at the time when this source is extinguished, 
and the remaining labels need to start competing for population. 

\begin{figure}
\begin{center}
\epsfig{figure=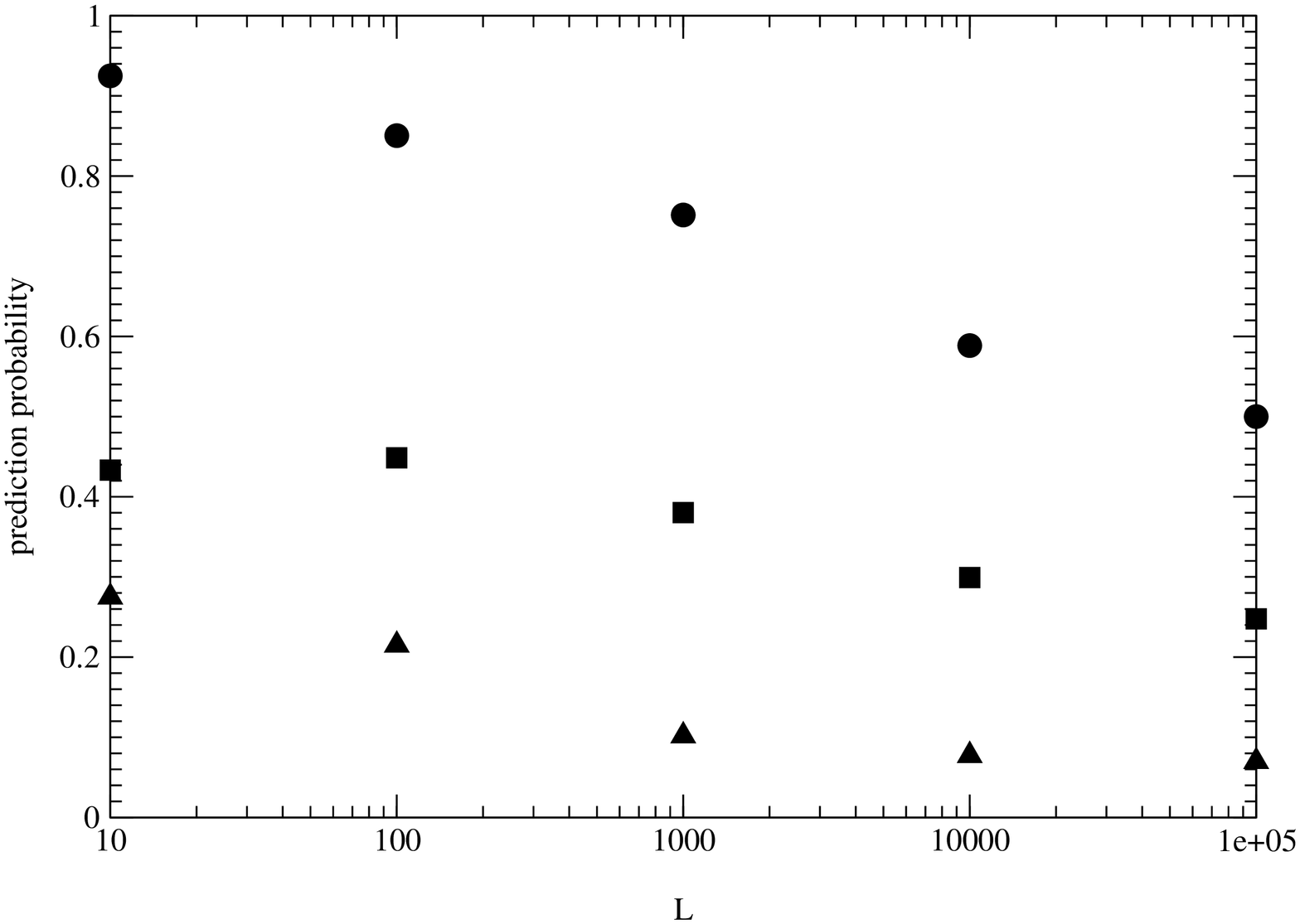,width=3.6in,angle=270}
\caption{The probability that the largest secondary cluster (squares), 
the fastest cluster (triangles) or the cluster with the highest product of 
momentum and size (circles) at the particular time when a dominant 
trend disappears becomes the new dominant trend, for a population 
with $N=10^3$, $p_{\rm crit}=10^{-5}$  and varying number of labels. 
As the number of labels increases calling the next winning trend 
with confidence becomes nearly impossible by any of these criteria.}  
\label{fig8}
\end{center}
\end{figure}

Is it possible, at this particular, time to predict the next winning trend?
There is a delicate balance between the visibility of a trend and its 
momentum. For example it is not true that the fastest growing  
trend becomes dominant and neither does the largest (secondary) cluster. 
In most cases the fastest moving trends are the smallest, which suffer 
from lack of visibility (i.e. it is unlikely to find an agent 
belonging to it). On the other hand a secondary large trend will 
not become dominant if the result of many inquires find it the laggard. 
Thus becoming the next winner requires being both relatively large,
being fast and some good portion of luck. Fig.~\ref{fig8} shows how well 
several of these criteria fare at predicting the winner. 

The results of Fig.~\ref{fig8} show that even if one is aware 
that a whole population is following trends it is difficult to call a 
winner early enough to profit from the realization. 
In fact, the most successful criterion shown in fig.~\ref{fig8},  
picking the trend with the largest product of size and momentum 
at the critical time, offers odds comparable to tossing a coin 
when $L$ becomes large. Moreover this strategy requires precise 
knowledge of the state of the whole population at a very particular time, 
which is in general difficult (and expensive) to obtain.

\section{Time averages and distributions: 
analogies to critical phenomena}
\label{sec4}

I have up to this point analyzed some of the general 
dynamical properties of the model introduced in sec.~\ref{sec2}. 
As we have seen there are no static stable solutions, and the 
population cycles between ordered states, where there is one dominant 
trend, and periods where many similar sized trends coexist and compete 
for dominance. In this section I discuss some of the time-averaged 
properties of these distributions, over many cycles of growth and collapse. 
Much like microcanonical time averages can coincide with canonical ensemble 
expectation values, time averages of complex systems give us a statistical 
measure of their most likely states. 
It is usually these averages that are compared to analogous quantities 
(over ensembles) in critical phenomena in order to demonstrate that the 
system displays self-organized criticality \cite{Bak}.
\begin{figure}
\begin{center}
\epsfig{figure=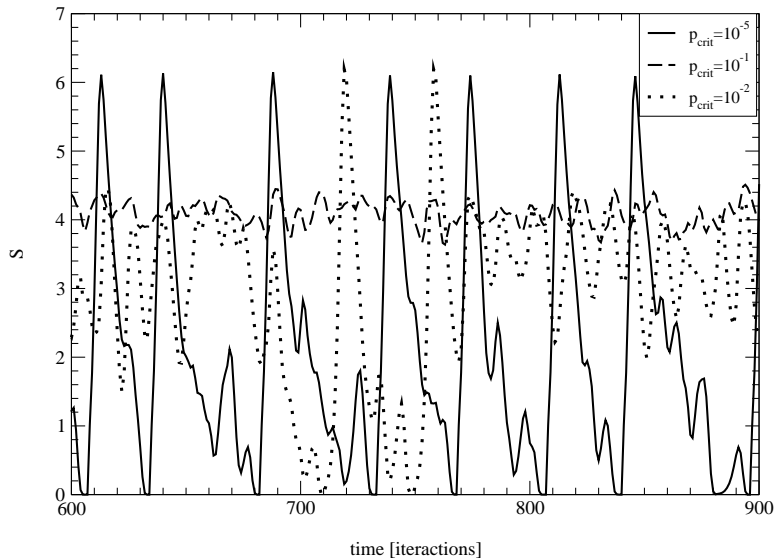,width=3.6in,angle=270}
\caption{The dependence of the time evolution of the total entropy 
$S$ on $p_{\rm crit}$ the conformist parameter for $N=10^5$ and 
$L=10^3$. $p_{\rm crit}$ controls
the amplitude and regularity of the trend cycling. 
For low enough $p_{\rm crit}\sim 1/N$ the motion is clearly cyclic alternating 
totally dominant trends with quasi-random distributions. 
As $p_{\rm crit}$ is increased the extremes  
become less pronounced and the motion less cyclic.}  
\label{fig9}
\end{center}
\end{figure}

Before analyzing time averages of distributions it is important to 
understand better another role of the conformism parameter $p_{\rm crit}$.
Fig.~\ref{fig9} shows how  $p_{\rm crit}$ determines the amplitude 
of the motion between order and disorder. Evolutions with low values 
of $p_{\rm crit} \leq N^{-1}$ reach  
absolute order (where one single trend engulfs the whole population) 
alternating with very disordered states. As $p_{\rm crit}$ is increased 
the amplitude decreases and the system spends more 
and more time in intermediate states. These are neither extremely 
ordered nor disordered but rather somewhere in between, as can be seen 
in an entropy plot, Fig.~\ref{fig9}.
\begin{figure}
\begin{center}
\epsfig{figure=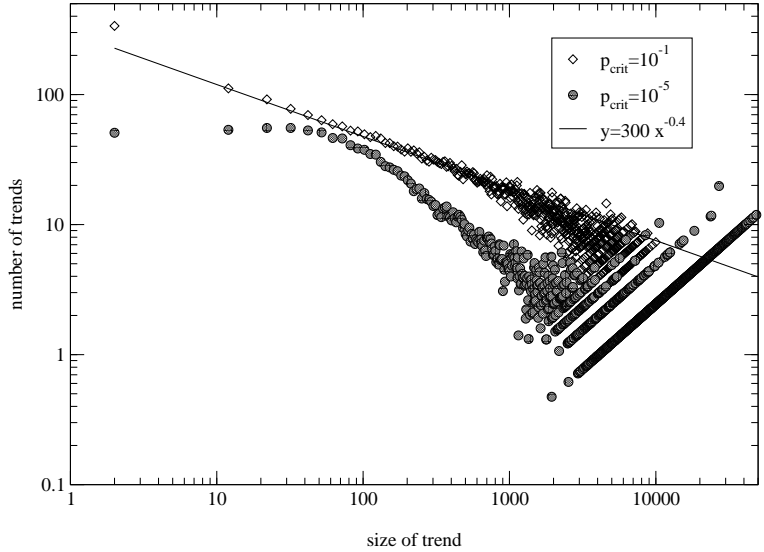,width=3.6in,angle=270}
\caption{The time-averaged trend size distribution for $N=10^5$, 
$L=10^3$ and two values of $p_{\rm crit}=10^{-1}, \ 10^{-5}$. 
$p_{\rm crit}$ controls the relative times spent by the system in each 
extreme configuration (trend dominance or disorder). 
For high enough values of $p_{\rm crit}$ 
the distribution becomes a power law with a small exponent, 
here $\alpha\sim 0.4$. For small $p_{\rm crit}$ the distribution 
is distorted by an abundance of both small and large trends.}  
\label{fig10}
\end{center}
\end{figure}

From these considerations about the role of $p_{\rm crit}$ we expect 
that time averages of distributions with small $p_{\rm crit}$ will 
include regions of configuration space with both more ordered and disordered 
states, those with higher $p_{\rm crit}$ will be peaked in between. 
This potentially leads to different time averaged distributions, 
as seen in Fig.~\ref{fig10}, which shows the number distribution 
of trend sizes.   
\begin{figure}
\begin{center}
\epsfig{figure=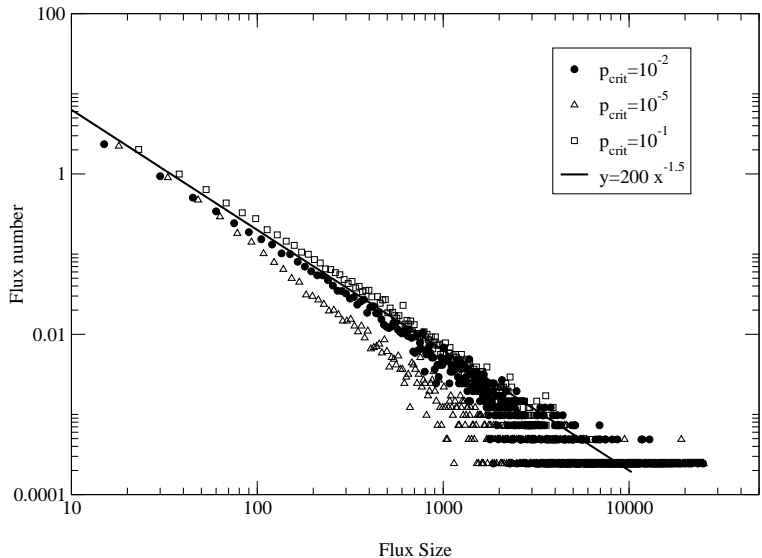,width=3.6in,angle=270}
\caption{The size distribution of population fluxes for the example
of Fig.~\ref{fig10}. 
The flux distributions shows significantly more robust power law behavior 
against variation of $p_{\rm crit}$ than the size distribution of trends.}  
\label{fig11}
\end{center}
\end{figure}

Another interesting quantity is the size distribution of population 
fluxes, defined as the number of individuals leaving 
or joining a trend per unit time. This is shown in Fig.~\ref{fig11}. 
The flux size distribution is essentially a dynamical process and shows 
much more robustness against changes of $p_{\rm crit}$.
A related quantity is the number distribution of individuals 
entering or leaving a trend after a test agent. 
\begin{figure}
\begin{center}
\epsfig{figure=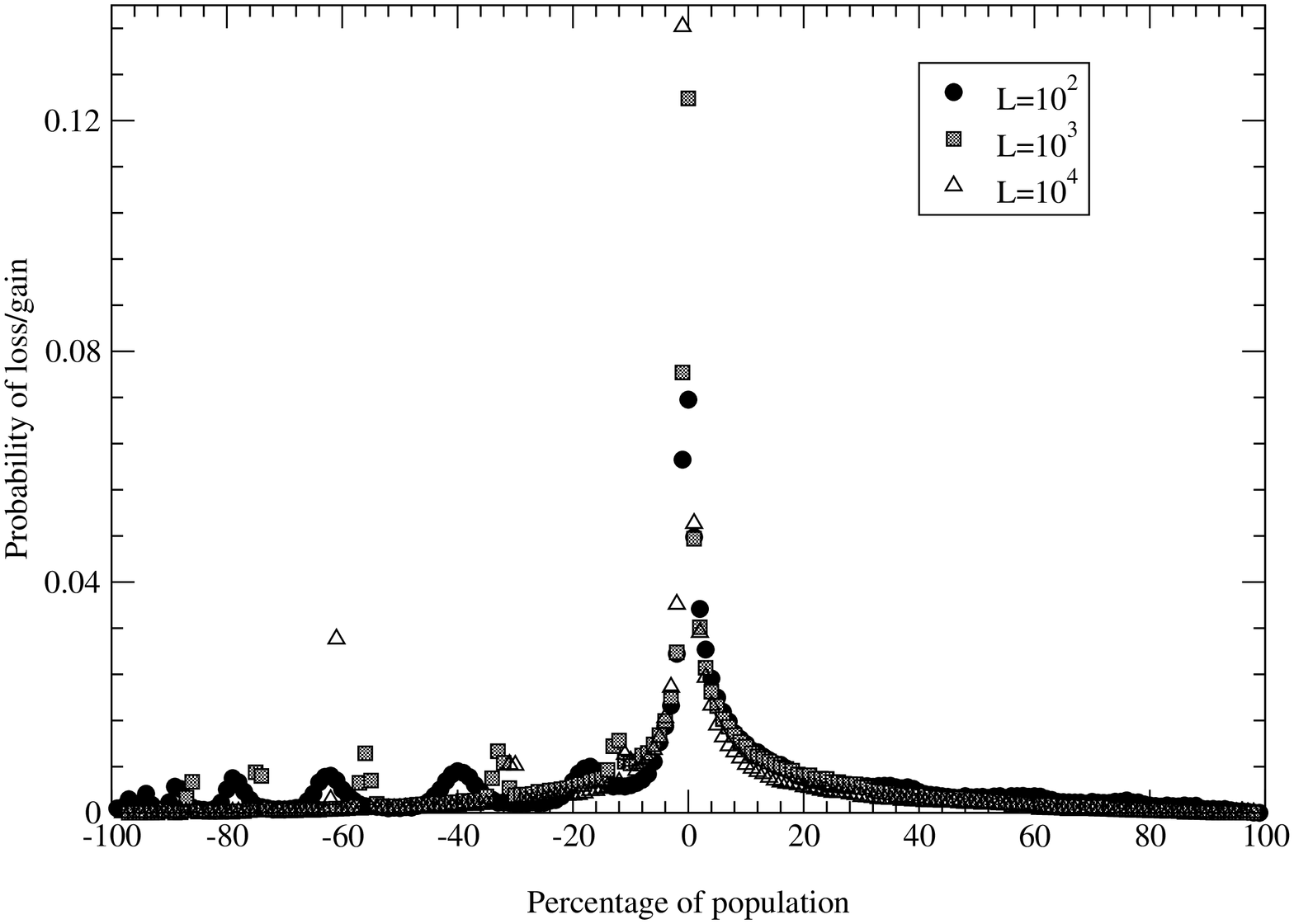,width=3.6in,angle=270}
\caption{The number distribution of individuals entering (positive) 
and leaving (negative) a trend after an agent has joined it for 
$N=10^4$, $p_{\rm crit}=10^{-5}$ and several $L$. 
On average an agent is followed by as many agents as those he follows. 
See also Fig.~\ref{fig13}. }  
\label{fig12}
\end{center}
\end{figure}

The number of individuals that join or leave after a given agent is a 
quantity that carries an important meaning. Think again of trend dynamics 
as a race, a game among agents. Ideally an agent wants
to join a winning trend early in its development and ride it until it is 
dominant; he also wants to leave before it starts decaying and start 
off the next winner. 
\begin{figure}
\begin{center}
\epsfig{figure=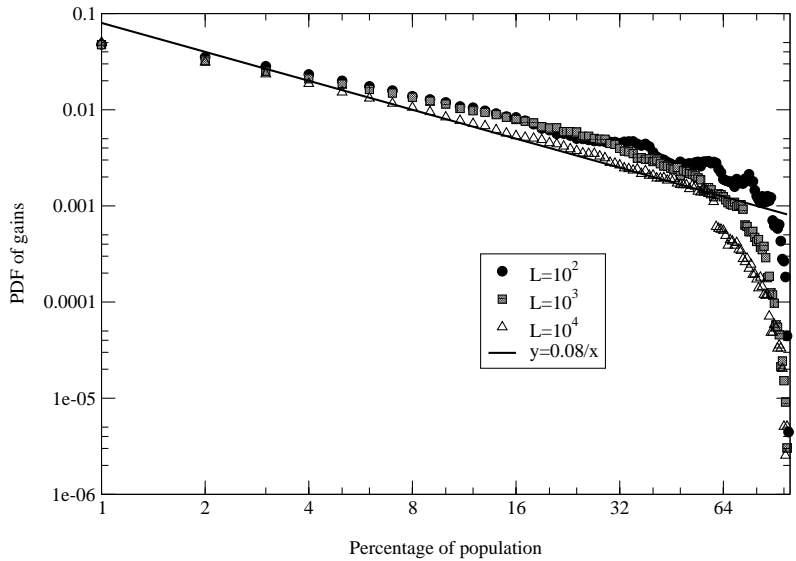,width=3.6in,angle=270}
\caption{Detail of Fig.~\ref{fig12} for agents entering a trend. 
The tail of the distribution is fat, leading to large higher moments. 
The distribution is eventually cutoff for large numbers 
by finite $N$ effects.}  
\label{fig13}
\end{center}
\end{figure}

This strategy 
is simply that of being followed by the maximum number of others and follow 
the least. This is desirable in many circumstances e.g. in 
(speculative) financial markets. Here, however, the cost of joining a trend 
does not increase with the number of agents already in it. 
Thus the present model can only hope to describe speculative bubbles in 
financial markets as long as price is no object for most agents.

Fig.~\ref{fig12} shows that over many cycles of trend growth and decay 
an individual is on average followed by as many others as those 
he follows. More interestingly the second moment of 
the distribution measures the possible fluctuations around zero 
gain or loss. If the game of following trends is played only a few times it 
will give a measure of the possible losses or gains incurred by a typical 
agent. Fig.~\ref{fig13} shows a detail of the distribution of gains. 
Although displaying finite population effects for very large numbers 
it is clear that the main distribution is very flat, making it 
possible for an individual playing the game only a few times to experience
spectacular gains and/or losses.

\section{Trend dynamics of open systems: changing populations and 
external information biases}
\label{sec5} 

So far I have considered the dynamics of trends over closed 
populations, characterized by a fixed number of agents $N$ and trends $L$. 
It is interesting to generalize this closed system to an open one, 
and examine the effects of adding or subtracting new individuals with 
particular preferences. This is the subject of this section.
\begin{figure}
\begin{center}
\epsfig{figure=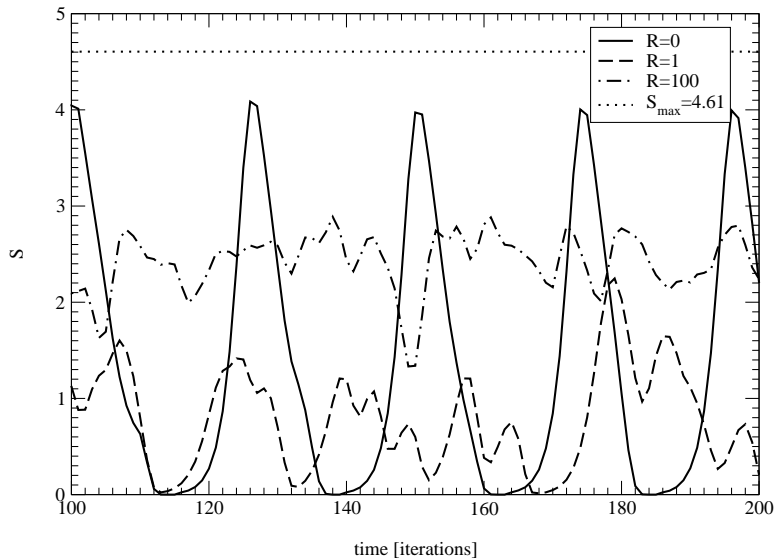,width=3.6in,angle=270}
\caption{The evolution of the total entropy $S$ for growing 
populations at different rates $R$ for $N(t=0)=5\times 10^4$ and 
$p_{\rm crit} =10^{-5}$. New agents enter the population 
at labels chosen at random. This leads  to the disappearance of 
the cyclic nature of the motion and in general to higher entropy states, 
particularly for large $R$.}  
\label{fig14}
\end{center}
\end{figure}

For purposes of illustration I consider a linear population growth law, i.e.
\begin{eqnarray}
N(t+1) = N(t) +R,
\label{Rate}
\end{eqnarray}
where $R$ is the number of individuals joining the populations at each time 
step. The exact form of Eq.~(\ref{Rate}) is not particularly 
important for our discussion.

The influence of the new individuals on a population undergoing trend 
dynamics depends sensitively on their trend preferences. 
First I investigate the effect of letting the new agents enter the 
population at a random trend, an "open minded youth". 
Note that this type of driving force, once time averaged, 
does not break the choice symmetry  
of the closed system (any trend may still become dominant). 
It does however introduce a {\it disordering} external effect 
on the dynamics, much like driving a statistical system with white noise. 

Fig.~\ref{fig14} shows the evolution of the total entropy 
under these circumstances, for several values of $R$.
The introduction of new agents at random, even in 
a population with small $p_{\rm crit}=10^{-5}$ and at small rates, 
leads to dramatic results. The distinctive cycling between order and 
disorder becomes less well defined even for the smallest $R=1$. 
The lifetime of the largest trends is reduced as they suffer from stiffer 
competition from smaller movements. These receive the strongest enhancement of 
their relative numbers and momentum due to the injection of new agents. 
Curiously the highest entropy states of the population are also suppressed, 
at least for small $R$. These states were formerly produced in the early 
decay of a very dominant trend. 
For large $R$ the average state of the population remains fairly 
disordered, but still far from the random distribution, which is presumably 
reached as $R\rightarrow \infty$.

\begin{figure}
\begin{center}
\epsfig{figure=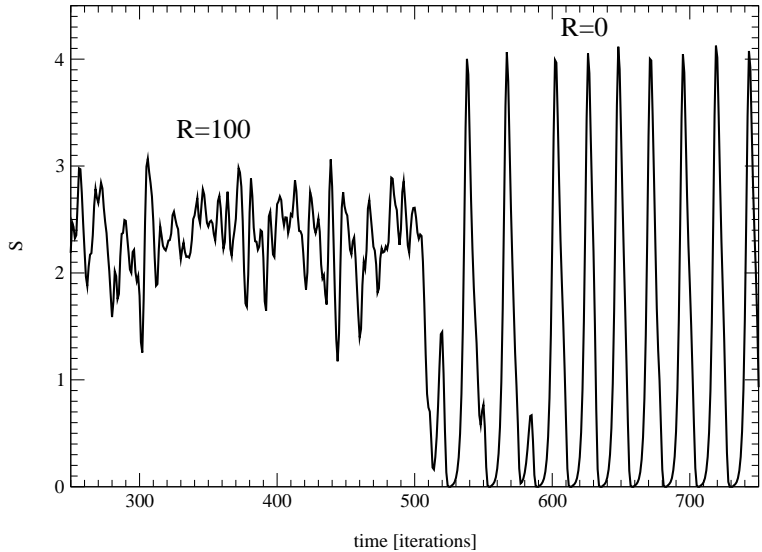,width=3.6in,angle=270}
\caption{The evolution of the total entropy for the growing population 
of Fig.~\ref{fig14} with $R=100$. 
The motion returns to its cyclic characteristic pattern once the 
inflow of new agents ceases (here at $t=500$).}  
\label{fig15}
\end{center}
\end{figure}

If after a period of growth the population ceases to increase, 
i.e. if the drive is switched off at some late time,  
the dynamics quickly resumes its cyclic pattern of dominant trend
 formation and decay, see Fig~\ref{fig15}, 
although now involving many more agents

Alternatively new agents may not be
introduced at random. This constitutes a systematic ordering effect 
that explicitly breaks trend choice symmetry. 
Because the evolution is characterized by a sequence of instabilities 
the dynamics are extremely sensitive to the preferences of the new agents. 
If new agents are introduced in a specific label then it becomes invariably 
the dominant trend, even at the smallest $R=1$. Similarly  if the 
new agents prefer a {\it subset} of the total labels, these become all the 
dominant trends. Remaining trends  gather at best a small fraction 
of the population and only at times when a dominant trend decays.

Finally I consider the effects of a decreasing population. The simplest 
and most natural situation is that individuals are eliminated from the 
population at random, according to Eq.~(\ref{Rate}), now with $R < 0$.
This operation actually has no effect on most aspects of the dynamics. 
This is because the change in momentum of each trend due to 
population losses is on average proportional to trend size 
(the probability that the individual 
is in that trend). Thus the change in momentum of a trend $i$ due to 
random population loss is $\Delta p_i = R n_i/(N n_i )= R/N$,  
the same for all trends. We conclude (and observe explicitly in 
the dynamics) that eliminating individuals at random preserves
the momentum hierarchy among trends and consequently does not change 
the dynamics. 

The exception to this rule is the decay of the single dominant trend.  
I postulated that only if the trend momentum is slow enough (but still 
positive) will agents 
leave it to start something new. 
However, due to population loss, the largest
trend may never come sufficiently close to $p=0$ (it would 
in a continuous time approximation). Then it is possible that all  remaining
agents become stuck in a single decaying trend. 
This happens at a characteristic time $t\simeq N/R - p_{\rm crit}^{-1}$,
obtained from comparing the momentum due to population losses alone to 
$p_{\rm crit}$, starting at $t=0$ with $N$ agents. 
For small $p_{\rm crit}$ the evolution freezes as the first large trend 
is formed. 

We see that random population growth or decay have 
quite different consequences. A growing population
with no particular preference at birth  makes spontaneous 
consensus rarer, more unstable and encompassing fewer agents.
Population growth with particular preferences, on the other hand, 
explicitly breaks the spontaneous choice symmetry of the closed model 
and reduces the space of possible large trends to those  preferred 
by the new agents. 
To combat this tendency the existing (older) elements of 
the population must not follow trends. 
Finally the effects of population decrease are generally 
more benign.  The important exception is that if a consensus occurs 
at a time of population decrease in a conformist population agents may become
stuck in a sinking trend. The same effect can take place if conformism 
rose with population loss and would have devastating consequences 
if diversity were necessary for eventual population recovery.

\section{Conclusions and outlook}
\label{sec6}

In this paper I introduced a simple agent based model describing how coherent 
social choices - trends - arise spontaneously and how 
these movements eventually slow down and disappear. 
Trends, fashions, fads are all well documented mechanisms
for the creation of shared knowledge (culture) and triggers for 
behavior change and its self-regulation in human societies. 
Examples range  from the most mundane (fads in popular culture) to the most 
important (setting cultural norms, fraud/crime control).

The dynamical pattern of trend formation and decay developed in this paper 
is particularly clear in populations with a high degree of conformism 
(small $p_{\rm crit}$), where individuals are slow to leave sluggish 
trends in order to start something new. 
Then the population goes through cycles alternating states of order, 
dominated by a single label, and disorder where many small trends compete 
for population. If conformism is lower the size and 
frequency of occurrence of dominant trends becomes smaller. As a result 
the state of the population remains somewhere halfway in between order and 
disorder and is characterized by scale invariant (power law) 
distributions - the dynamics oscillates about a state 
of criticality. The amplitude of fluctuations around this state is also
determined by the conformism parameter $p_{\rm crit}$. The passage 
through criticality is made apparent by the construction 
of familiar quantities from statistical mechanics, analogous to percolation 
cumulants. 

The fundamental instabilities characteristic of both asymptotic states 
of absolute order and absolute disorder make the dynamics extremely 
susceptible to external influences, such as the preferences of new 
individuals. Simple as the model is it suggests that variety can be sustained 
in an open population following trends only if new agents (youth) 
remain unprejudiced. 
On the other hand diversity of choices will be quenched if their attention 
spans only a subset of all pre-existing possibilities. 
To prevent this loss of diversity it is then necessary that older agents 
stick to their own preferences and do not follow trends.  Population loss 
at random on the other hand does not affect most aspects of trend dynamics 
but may lead to a scenario where the state of the population is frozen 
in a single decaying trend if conformism is too high. The maintenance 
of diversity in this situation requires that agents continue to take 
risks in the face of adversity. 

The spontaneous breaking of choice symmetry inherent to trend 
dynamics illustrates a simple but important point about decision making 
strategies. In the present model there is no extremum principle 
that makes the choice of a winning trend predictable. Instead a consensus 
is built out of accidental elements of the dynamics and, paradoxically, 
the desire of individual agents to stay ahead of the crowd. 
Thus a trend that finds itself 
large and fast growing at the right moment will have a sort of 
first striker's advantage. This leads to accidental winners that may not 
be optimal solutions in the long run, a phenomenon familiar to technological
and (presumably) biological development.  

The most important consequence of the difficulty 
to predict winning trends is that, although it may be apparent to an 
external observer that everyone is following trends, it is extremely 
difficult to take advantage of the phenomenon for individual profit.
This interesting aspect of the problem will be discussed further in a 
future publication \cite{bett2}. There I will consider how trend dynamics 
can be formulated as a game played for social advantage over a typical 
social network. It will then become apparent that some of morphological 
characteristics  of social networks are not accidental, but rather 
may follow from our collective desire to access socially relevant 
information as early as possible and from the necessity to 
keep up with our friends and neighbors.

\section*{Acknowledgments} 

I am much indebted to Stirling Colgate for awakening my 
interest in several questions in complex systems. 
I would also like to thank Christoph Best for several 
discussions about the workings of the model and diagnostics and  
Nuno Antunes, Yoav Bergner, Luis Rocha and Ben Svetitsky 
for useful comments.


\begin{thebibliography}{99}


\bibitem{Keynes} In Economics the importance of trends was probably 
first expressed in print by  J.~M.~Keynes, {\it The general theory of Employment, 
Interest and Money} (Macmillan, London, 1936).

\bibitem{Social} See e.g. the discussion and references in 
C.~Cornell and S.~Cohn, Am. J. Sociol. {\bf 101}, 366 (1995). 

\bibitem{hirshleifer} S. Bikhchandani, D. Hirshleifer and I. Welch,
J. Polit. Econ. {\bf 100}, 992-1026 (1992); 
J. Econ. Perspect. {\bf 12}, 151 (1998).

\bibitem{BHWrefs} The reader is encouraged to pursue 
the few hundred publications that cite Ref.~\cite{hirshleifer} 
across several fields. For a few highlights taken from this set and beyond 
see \cite{Social,Analysts,FundManagers,Voting,IndustrialInnovation,TVfads,Crime}.

\bibitem{earlyherd}
For early arguments and analysis see K.~D.~West, 
J. Financ. {\bf 43}, 639 (1988) and references therein. 
For early models of herding in financial markets see
R.~Topol, Econ. J. {\bf 101}, 768 (1991);
A.~Bannerjee, Q. J. Econ. {\bf 107}, 797 (1992); 
Rev. Econ. Stud. {\bf 60}, 309 (1993);
A.~Orl\'ean, J Econ. Behav. Organ. {\bf 28}, 257 (1995); 
T.~Lux, Econ. J. {\bf 105}, 881 (1995).


\bibitem{Analysts}  
B.~Trueman, Rev. Financ. Stud. {\bf 7}, 97 (1994); 
I.~Welch, J. Financ. Econ. {\bf 58}, 369 (2000); 
H.~Rao, H.~R.~Greve, G.~F.~Davis,
Admin. Sci. Quart. {\bf 46}, 502 (2001). 

\bibitem{FundManagers}  
D.~S.~Sharfstein and J.~C.~Stein, Am. Econ. Rev. {\bf 80}, 
465 (1990);  M.~Grinblatt, S.~Titman, and R.~Wermers, 
Am. Econ. Rev. {\bf 85}, 1088 (1995).   


\bibitem{Voting} 
R.~D.~McKelvey and P.~C.~Ordeshook, J. Econ. Theory {\bf 36}, 55 (1985); 
L.~M.~Bartels, {\it Presidential Primaries and the Dynamics of 
Public Choice} (Princeton Univ. Press, Princeton NJ, 1988); 
A.~Cukierman, {\it Asymmetric Information and the Electoral 
Momentum of Public Opinion Polls} (Princeton University Press,
Princeton, 1989) and references therein. 


\bibitem{IndustrialInnovation} 
R.~J.~Gilbert and M.~Lieberman, Rand J. Econ. {\bf 18}, 17 (1987);  
W.~Pesendorfer, Am. Econ. Rev. {\bf 85}, 771 (1995). 

\bibitem{TVfads}  R.~E.~Kennedy,
J. Ind. Econ. {\bf 50}, 57 (2002). 

\bibitem{Crime}
S.~M.~Sheffrin and R.~K.~Triest, in {\it Why People pay taxes} 
Ed. J.~Slemrod (1992); E.~Glaeser, B.~Sacerdote and J.~Scheikman, 
Q. J. Econ. {\bf 111}, 507 (1996).

\bibitem{StatPhys} For other models of social decision making 
inspired by concepts of Statistical Physics see 
D.~Chowdhury, D.~Stauffer, E. Phys. J. B {\bf 8}, 477 (1999);
D.~Stauffer, N.~Jan, Int. J. Mod. Phys. C {\bf 11}, 147 (2000);
K.~Sznajd-Weron, J.~Sznajd, 
Int. J. Mod. Phys. C {\bf 11}, 1157 (2000);
D.~Stauffer, A.~O.~Sousa, S.~M.~de~Oliveira,
Int J. Mod. Phys. C {\bf 11}, 1239 (2000);
R.~Florian, S.~Galam, E. Phys. J. B {\bf 16}, 189 (2000);
S.~Solomon, G.~Weisbuch, L.~de~Arcangelis, N.~Jan, D.~Stauffer, 
Physica A {\bf 277}, 239 (2000).

\bibitem{sneppen} 
R.~Donangelo, K.~Sneppen, Physica A  {\bf 276}, 572 (2000);
R.~Donangelo, A.~Hansen, K.~Sneppen, S.~R.~Souza, Physica A  {\bf 287}, 
539 (2000).

\bibitem{flocks} 
J.~Toner, Y.~Tu, Phys. Rev. Lett.  {\bf 75}, 4326 (1995);
{\it ibid.} {\bf 80}, 4819 (1998); Phys. Rev. E {\bf 58}, 4828 (1998);
A.~S.~Mikhailov and D.~H.~Zanette, Phys. Rev. E {\bf 60}, 4571 (1999);
A.~Czir\'ok, A.-L.~Barab\'asi, and T.~Vicsek,
Phys. Rev. Lett. {\bf 82}, 209 (1999).

\bibitem{herd} More recent models of herd behavior in financial markets 
inspired by statistical physics are 
R.~Cont and J.-P.~Bouchaud, Macroecon. Dyn. {\bf 4}, 170 (2000);
T.~Lux and M.~Marchesi, Nature (London) {\bf 397}, 498 (1999);
V.~M.~Eguiluz and M.~G.~Zimmermann, Phys. Rev. Lett. {\bf 85}, 5659 (2000).

\bibitem{herd2} For more extensive analysis and further 
exploration of the class of models introduced in \cite{herd} see 
D.-F.~Zheng, G.~J.~Rodgers, P.~M.~Hui, R.~D'Hulst,
Physica A {\bf 303}, 176 (2002); 
G.~J~Rodgers, D.-F. Zheng,  Physica A {\bf 308}, 375 (2002); 
D.-F.~Zheng, P.~M.~Hui, K.~F.~Yip, N.~F.~Johnson, 
E. Phys. J. B {\bf 27}, 213  (2002); F.~Wagner, cond-mat/0207280.

\bibitem{Bak} 
P.~Bak, C.~Tang, and K.~Wiesenfeld, Phys. Rev. Lett. {\bf 59},  381 (1987); 
see also P.~Bak, {\it How nature works: the science of self-organized criticality} 
(Copernicus, New York, NY, USA 1996) and references therein.


\bibitem{percolation} See e.g. D.~Staufer and A.~Aharony, {\it Introduction 
to percolation theory} (Taylor \& Francis, London, UK, 1992); 
K.~Binder and D.~W.~Heermann, {\it Monte Carlo simulation in Statistical 
Physics} (Springer-Verlag, Heidelberg, Germany 1997). 

\bibitem{function} Clearly the same dynamics results from the 
comparison between agents of any growing function of their trend's 
momentum.

\bibitem{SocialNReview} For recent reviews of the structure of social 
networks see D.~J.~Watts, {\it Small worlds} (Princeton University 
Press, Princeton, 1999); M.~E.~J.~Newman,  J. Stat. Phys. {\bf 101}, 
819 (2000).


\bibitem{bett2} L.~M.~A. Bettencourt, {\it Trend dynamics  and the 
structure of social networks}, in preparation.

\bibitem{Movies} Animations of the evolution of the trend occupation 
numbers over many cycles of trend formation and decay 
can be seen at http://www.mit.edu/$^\sim$lmbett.



\end{thebibliography}
\end{document}